\documentclass[prd,aps,twocolumn,tighten]{revtex4}
\usepackage{latexsym}
\begin{document}
\newcommand{\beq}{\begin{equation}}
\newcommand{\eeq}{\end{equation}}
\newcommand{\Prd}{Phys. Rev D}
\newcommand{\Prl}{Phys. Rev. Lett.}
\newcommand{\Plb}{Phys. Lett. B}
\newcommand{\Cqg}{Class. Quantum Grav.}
\newcommand{\Np}{Nuc. Phys.}
\title{Geometrical description of spin-2 fields}
\author{M. Novello}
\address{\mbox{}\\Centro Brasileiro de Pesquisas F\'{\i}sicas,
 \\Rua Dr.\ Xavier Sigaud 150, Urca 22290-180, Rio de Janeiro, RJ -- Brazil
\\E-mail: novello@cbpf.br}
\date{\today}

\begin{abstract}
We show that the torsion of a Cartan geometry can be associated to
two spin-2 fields. This structure allows a new approach to deal
with the proposal of geometrization of  spin-2 fields besides the
traditional one dealt with in General Relativity.  We use the
associated Hilbert-Einstein Lagrangian $R$ for generating a
dynamics for the fields.
\end{abstract}

 \maketitle
\smallskip\mbox{}

\renewcommand{\thefootnote}{\arabic{footnote}}

\section{Introduction}

\subsection{Introductory remarks I}

From the  particle physicists point of view the use of a spin-2
field represented by a second order symmetrical tensor
$\varphi_{\mu\nu}$ in order to define a metric tensor $g_{\mu\nu}$
of a Riemannian spacetime seems quite natural. Both quantities
have the same tensorial character and the same explicit symmetry.
In other words, if one faces the question of how to associate a
symmetric second order tensor $\varphi_{\mu\nu}$ that represents a
spin-2 field, into a geometrical framework,  the lesson we learned
from last century leads almost univocally to a prompt answer: one
should add this field to the Minkowski metric tensor of the flat
background spacetime and generate a curved Riemannian geometry by
defining an associated metric tensor by the expression:
\begin{equation}
 g_{\mu\nu} \equiv \eta_{\mu\nu} + \varphi_{\mu\nu}.
 \label{001}
 \end{equation}
Note that this is not an approximation but an exact relation. To
obtain from the above expression the corresponding inverse
$g^{\mu\nu}$ defined by
$$g^{\mu\nu} \, g_{\nu\alpha} = \delta^{\mu}_{\alpha}$$
one is led to the infinite series
\begin{equation}
g^{\mu\nu} = \eta^{\mu\nu} - \varphi^{\mu\nu} +
\varphi_{\mu\alpha} \, \varphi^{\alpha\nu} +  \cdots
\label{002}
\end{equation}
 With these definitions it follows the inevitability of a nonlinear process to
implement the theory based on the identification equation
(\ref{001}). This kind of geometrical formulation is the basis of
Einstein General Relativity (GR).
Although  Einstein path starts from an a priori geometrical
description, there is an equivalent description that is based on a
field theoretical formulation. The best way to understand this is
to follow the work of Gupta, Deser, Feynmann and others. See
\cite{NDL} for details and further references.

A dynamics within this geometrical scheme is obtained from
Hilbert-Einstein action in terms of the scalar of curvature of the
associated Riemannian geometry:
\begin{equation}
S = \int \sqrt{-g} \, R \,d^{4}x.
\label{I2}
\end{equation}
The success of such procedure led to the belief that in
4-dimension there is no way to introduce any other spin-2 field in
the geometry. Or, at least, not in such simple and natural way,
as is done in the above definition (\ref{001}).

The purpose of this paper is precisely to discuss on this and to
point out the existence of another "natural" way to associate
spin-2 fields to the geometry. In other words we will show an
alternative approach to
geometrize a spin-2 field.
 Let me stress from the beginning that we will not deal here
 with an alternative way to describe gravitational
effects. Our work here is just to examine the geometrical
consequences of the existence of others spin-2 fields besides the
gravitational field which is associated to the metric tensor. The
motivation of this paper is precisely this: to report on a new
form to integrate spin-2 field in the geometry. We will show that
the scenario which allows for such task already exists and is
realized by means of a well known generalization of the riemannian
structure of spacetime just by assuming the presence of torsion in
the geometry. In other words, one has to consider Cartan geometry
\cite{Cartan}, which, besides the metric, contains an
antisymmetric torsion $\tau^{\alpha}_{\mu\nu}.$

In order to understand such identification of the spin-2 field
with torsion we have to start by recovering a property of field
theory established in the first part of the last century by Fierz.
The main result states that a spin-2 field can be described in two
equivalent ways, which we will call the Einstein-frame and the
Fierz-frame representations. The most common one, the
Einstein-frame, uses a symmetric second order tensor
$\varphi_{\mu\nu}$ to represent the field. In the Fierz-frame this
role is played by a third order tensor $F_{\alpha\mu\nu},$ which
is antisymmetric in the first pair of indices \cite{fierz}:
$F_{\mu\nu\alpha} = - F_{\nu\mu\alpha}.$
The existence of this Fierz-representation opens a new way to
associate spin-2 field to the geometry which seems as natural as
the one pointed out above. We shall see that this new form of
geometrization depends crucially on the uses of Fierz
representation of the spin-2 field in terms of a third rank
tensor. The oblivion of Fierz representation along the last fifty
years and the success of General Relativity were responsible for
the general belief that a symmetric second order tensor is the
unique way to represent a spin-2 field and consequently is the
reason for which such scheme of geometrization  was not noticed
before.

Our analysis rest on the simplest generalization of Riemann
geometry that deals with an affine connection which is not
symmetric. In other words, the connection is written as the sum of
the Christoffel symbol and an extra tensor defined by the torsion
tensor $\tau^{\alpha}_{\mu\nu}.$ This tensor is antisymmetric:
$\tau^{\alpha}_{\mu\nu} =  - \tau^{\alpha}_{\nu\mu}.$ We shall see
that as in the case of spin-2 description in terms of a symmetric
second order rank, in the case in which the space is endowed with
a torsion it will be equally "natural" to associate the torsion to
spin-2 tensors in the Fierz representation.

\subsection{Introductory Remarks II}

Recently \cite{novello/neves} we have analysed the formulation of
spin-2 field (massive and massless) in terms of a third order
tensor $F_{\alpha\mu\nu},$. As a consequence of such exam we
concluded that in flat Minkowski spacetime both variables are
equivalent: the dynamics is the same and the corresponding
structure of the consistency of the dynamical equations is
completely equivalent. Nevertheless, in the case of a curved
spacetime this is no longer true. We have shown that the use of
the Fierz-frame seems more compelling, once it yields, through the
standard minimal coupling principle, a unique, non-ambiguous
description. The use of the Einstein-frame in the passage from the
flat Minkowskii background to the curved Riemannian one in General
Relativity, introduces ambiguities which come from the non
commutativity of the covariant derivatives. In
\cite{novello/neves} we have shown that the use of the Fierz-frame
gives an unambiguous minimal coupling treatment for equations that
are equivalent to those studied by Aragone-Deser and Buchbinder et
al. Thus, not only the Fierz frame is equivalent to Einstein frame
to describe spin-2 fields but, more than this, it avoids the
arbitrariness and inconsistency that exists in the standard
formulation of a spin-2 field coupled to gravity in
Einstein-frame. Besides, the superiority of the Fierz frame
appears more explicitly in the combined set of equations for
spin-2 field and gravity: it preserves the standard Einstein
equations of motion, whilst maintaining the correct degrees of
freedom for the spin-2 field.

The idea that torsion can be associated to fundamental fields has
been examined previously (see for instance \cite{hammond}) in a
tentative to incorporate torsion as a manifestation of a
completely  antisymmetric field typical of string theories. In the
present paper we intend to present another way to look into this
problem by incorporating two spin-2 tensors into the Cartan
geometrical scheme.

In order to exhibit more clearly the properties of this
identification we start with the simplest geometrical structure
embodied with torsion and a Minkowski metric tensor
$\eta_{\mu\nu}.$  The dynamics mimics General Relativity using the
scalar of curvature as a natural generalization of
Hilbert-Einstein action. In this particular case the curvature is
a consequence only of the presence of the torsion part of the
affine connection, since the Riemannian part of the curvature
--which depends on the metric tensor and on the associated
Christoffel symbol -- vanishes.


\subsection{Synopsis}
In this paper we will deal with Cartan geometry. In section II we
present a set of definitions and equations which will be needed in
this paper. Section III deals with the particular case of
restricted Cartan geometry in which the degrees of freedom of the
torsion field are reduced from 24 to 10. We present the affine
connection and the associated curvature tensor. We use the scalar
of curvature of the restricted case in order to produce a dynamics
for torsion. We show that this dynamics coincides with the
standard dynamics for a spin-2 field, that is, the linearized
Einstein equation of GR. In the next section we generalize this
formalism and deal with 20 degrees of freedom. We use the same
structure of the scalar of curvature to produce a dynamics for
torsion and we show that it is nothing but the same dynamics for
two non-interacting spin-2 fields. In the last section we deal
with the interaction of these spin-2 fields with matter using the
minimal coupling principle. We end with some comments on this and
some perspectives for future work.

\section{Some mathematical machinery}

\subsection{Fierz representation of spin-2 field}

We define a three-index tensor $F_{\alpha\beta\mu}$ which is
anti-symmetric in the first pair of indices and obeys the cyclic
identity, that is

\begin{equation}
F_{\alpha\mu\nu} + F_{\mu\alpha\nu} = 0 \label{F01}
\end{equation}
\begin{equation}
F_{\alpha\mu\nu} + F_{\mu\nu\alpha} + F_{\nu\alpha\mu} = 0.
\label{F02}
\end{equation}

This last expression means that the dual of $F_{\alpha\mu\nu}$ is
trace-free:

\begin{equation}
\stackrel{*}{F}{}^{\alpha\mu}{}_{\mu} = 0 , \label{02bis}
\end{equation}
where the asterisk represents the dual operator, defined in terms
of the completelly anti-symmetric object
$\eta_{\alpha\beta\mu\nu}$ by
\[
\stackrel{*}{F}{}^{\alpha\mu}{}_{\lambda} \equiv \frac{1}{2} \,
\eta^{\alpha\mu}{}_{\nu\sigma}\,F^{\nu\sigma}{}_{\lambda}.
\]
The dual of the Levi-Civita object allows the introduction of the
tensor $g_{\alpha\beta\mu\nu}$ once we have the equality
\begin{equation}
g_{\alpha\beta\mu\nu} = - \,{\eta}^{*}_{\alpha\beta\mu\nu},
\label{03bis2}
\end{equation}
where
\begin{equation}
g_{\alpha\beta\mu\nu} \equiv g_{\alpha\mu}\,g_{\beta\nu} -
g_{\alpha\nu}\, g_{\beta\mu}.
 \label{03bis}
\end{equation}

\subsection{The case of a single spin-2 field}

The field $F_{\alpha\mu\nu}$ has 20 independent components. In
order to eliminate the extra $10$ independent components and allow
it to represent a single spin-2 field, we impose an additional
requirement contained in the following lemma\footnote{The proof of
this and subsequent lemmas are given in the quoted paper by
Novello et al.}: The necessary and sufficient condition for
$F_{\alpha\mu\nu}$ to represent an unique spin-2 field is

\begin{equation}
\stackrel{*}{F}{}^{\alpha (\mu\nu)}{}_{,\alpha} = 0.  \label{F03}
\end{equation}

We represent the symmetrization symbol by $A_{(\mu\nu)} \equiv
A_{\mu\nu}+A_{\nu\mu}.$ We use an analogous form for the
anti-symmetrization symbol: $[x, y] \equiv xy - yx.$

 We will call a tensor that satisfies conditions (\ref{F01}),
(\ref{F02}) and (\ref{F03}) a {\bf Fierz tensor}.

Condition (\ref{F03}) implies that there exists a symmetric second
order tensor $A_{\mu\nu} = A_{\nu\mu}$ such that we can write
\begin{equation}
2 F_{\alpha\mu\nu} = A_{\nu\alpha,\mu} - A_{\nu\mu ,\alpha} -
g_{\alpha\mu\nu\epsilon}  F^{\epsilon}.
  \label{06}
\end{equation}
The factor $2$ in the l.h.s. is introduced for convenience.

\begin{equation}
F_{\alpha} \equiv F_{\alpha\mu\nu} g^{\mu\nu} = A_{,\alpha} -
A_{\alpha}{}^{\lambda}{}_{,\lambda}, \label{05}
\end{equation}
and
 $ A \equiv A_{\mu\nu} g^{\mu\nu}.$

 When a Fierz tensor is written under the form given
 in equation (\ref{06}) we will say that it is in the Einstein
frame. This formula can be made covariant and generalized
trivially for arbitrary system of coordinates.

A Fierz tensor $F_{\alpha \mu \nu }$ satisfies the identity
\begin{equation}
F^{\alpha }{}_{(\mu \nu ),\alpha }\equiv -\,G^{L}{}_{\mu \nu } ,
\label{07}
\end{equation}
where $G^{L}{}_{\mu \nu }$ is the linearized Einstein operator
defined in terms of the symmetric tensor $A _{\mu \nu }$ by

\begin{equation}
G^{L}{}_{\mu \nu }\equiv \Box \,A _{\mu \nu } - A^{\epsilon
}{}_{(\mu ,\nu )\,,\epsilon }+ A_{,\mu \nu }-g _{\mu \nu }\,\left(
\Box A -A^{\alpha \beta }{}_{,\alpha \beta }\right) . \label{08}
\end{equation}
\subsection{Dynamics}

We limit all our considerations in the present paper to a dynamics
for the Fierz field which is linear. The most general linear
theory comes from a combination of the invariants one can
construct with the field. There are two\footnote{there is another
one $Z \equiv F_{\alpha \beta
\lambda}\stackrel{\ast{F}}{}^{\alpha\beta\lambda}$ which we will
not consider here once it is a topological invariant.} of them
which we represent by $X$ and $Y:$

\begin{eqnarray}
X &\equiv &F_{\alpha \mu \nu }\hspace{0.5mm}F^{\alpha \mu \nu }  \nonumber \\
Y &\equiv &F_{\mu }\hspace{0.5mm}F^{\mu.}  \nonumber \\
\label{AB}
\end{eqnarray}
The standard equation for the massless spin-2 field, the
linearization of Einstein equations $R_{\mu\nu} - \frac{1}{2} R
g_{\mu\nu}$ is given by
\begin{equation}
G^{L}{}_{\mu \nu }=0,  \label{014bis}
\end{equation}
or, in an equivalent way, using the identity (\ref{07})
\begin{equation}
F^{\lambda (\mu \nu )}{}_{,\lambda }=0.  \label{014}
\end{equation}
The corresponding action takes the form
\begin{equation}
S=\frac{1}{k}\,\int (X-Y) \, {\rm d}^{4}x. \label{013}
\end{equation}
Note that the Fierz tensor has dimensionality (lenght)$^{-1}.$
Thus the constant $k$ has dimensionality (energy)$^{-1}$
(lenght)$^{-1}$.

\subsection{The case of two spin-2 fields}
In the general case, in which condition (\ref{F03}) is not
satisfied, $F^{\lambda \mu \nu}$ represents two spin-2 fields.  In
this case $F_{\epsilon\nu\mu}$  has 20 independent components. We
use the Fierz decomposition in terms of two symmetric second order
tensors $A_{\mu\nu}$ and $B_{\mu\nu}$ that is:

\begin{equation}
F_{\alpha\beta\mu} = A_{\alpha\beta\mu} + \frac{1}{2}
\,\eta_{\alpha\beta}\mbox{}^{\rho\sigma} \, B_{\rho\sigma\mu},
\label{gc6}
\end{equation}
where
\begin{equation}
2 \,A_{\alpha\beta\mu} \equiv A_{\mu[\alpha,\beta]} -
g_{\alpha\beta\mu\epsilon} \,A^{\epsilon} \label{gc7}
\end{equation}
\begin{equation}
2 \, B_{\alpha\beta\mu} \equiv B_{\mu[\alpha,\beta]} -
g_{\alpha\beta\mu\epsilon} \,B^{\epsilon} \label{gc8}
\end{equation}
\begin{equation}
A_{\lambda} \equiv A_{,\lambda} -
A_{\lambda}\mbox{}^{\mu}\mbox{}_{,\mu} \label{gc8bis}
\end{equation}
\begin{equation}
B_{\lambda} \equiv B_{,\lambda} -
B_{\lambda}\mbox{}^{\mu}\mbox{}_{,\mu}, \label{gcbis2}
\end{equation}
with $A \equiv A_{\mu\nu}\,g^{\mu\nu}$ and $B \equiv
B_{\mu\nu}\,g^{\mu\nu}.$  Note that the trace and the pseudo-trace
of the tensor $F_{\alpha\beta\mu}$ are, respectively,
\begin{equation}
F_{\alpha\beta}\mbox{}^{\beta} = A_{\alpha} \label{gc9}
\end{equation}
\begin{equation}
F^{*}_{\alpha\beta}\mbox{}^{\beta} = -\,B_{\alpha}. \label{gc10}
\end{equation}

 \subsection{Torsion}

Let us now consider a four dimensional spacetime endowed with a
metric tensor $g_{\mu\nu}$ and a non symmetric affine connection
$\Gamma^{\alpha}_{\mu\nu}$ that defines a covariant derivative.
This structure is usually called a Cartan space. For an arbitrary
vector $V^{\mu},$ the covariant derivative is defined by:
\begin{equation}
V^{\mu}\mbox{}_{;\nu} \equiv V^{\mu}\mbox{}_{,\nu} +
\Gamma^{\mu}_{\nu\alpha} V^{\alpha}. \label{01}
\end{equation}
The torsion tensor $\tau^{\alpha}_{\mu\nu}$ is given by
\begin{equation}
\tau^{\alpha}_{\mu\nu} \equiv \frac{1}{2} \,
\left(\Gamma^{\alpha}_{\mu\nu} - \Gamma^{\alpha}_{\nu\mu} \right).
\label{02}
\end{equation}
It is worthwhile to decompose torsion tensor into its irreducible
components by setting:
\begin{equation}
\tau^{\alpha}_{\mu\nu} = L^{\alpha}_{\mu\nu} + \frac{1}{3} \,
\left(\delta^{\alpha}_{\mu}\, \tau_{\nu} - \delta^{\alpha}_{\nu}\,
\tau_{\mu}\right) - \frac{1}{3} \,
\eta_{\mu\nu}\mbox{}^{\alpha\lambda}\,  \tau^{*}_{\lambda}
\label{03}
\end{equation}

The quantity $\tau_{\mu} = \tau^{\alpha}\mbox{}_{\alpha\mu}$ is
the trace while $\tau^{*}_{\mu} =
\tau^{\alpha}\mbox{}^{*}_{\alpha\mu}$ is the pseudo-trace. Note
that the condition of metricity is satisfied, that is
\begin{equation}
g_{\mu\nu ;\alpha} = 0. \label{04}
\end{equation}

\section{Restricted Cartan Geometry (RCG)}

In this section we will deal with the simple case of a
pseudo-traceless Cartan geometry, that is we set:
\begin{equation}
\tau^{*}_{\alpha} = 0. \label{0001}
\end{equation}
This condition diminishes the 24 degrees of freedom of torsion to 20.

It is useful to represent torsion by an associated
quantity $F_{\mu\nu}\mbox{}^{\alpha}$ defined by the combination
\begin{equation}
\tau^{\alpha}\mbox{}_{\mu\nu} -
g^{\alpha\epsilon}\mbox{}_{\mu\nu}\,\tau_{\epsilon} =
F_{\mu\nu}\mbox{}^{\alpha}. \label{0002}
\end{equation}
It follows that this tensor $F_{\mu\nu\alpha}$ is
anti-symmetric in the first two indices:
\begin{equation}
F_{\mu\nu\alpha} = - F_{\nu\mu\alpha}. \label{0003}
\end{equation}
Besides, since the torsion tensor has no pseudo-trace it follows,
from definition (\ref{0002}) the additional symmetry
 \begin{equation}
F_{\mu\nu\alpha} + F_{\nu\alpha\mu} + F_{\alpha\mu\nu} = 0.
\label{0005}
\end{equation}
The tensor $F_{\mu\nu\alpha}$ so defined has, like the torsion
with vanishing pseudo trace, 20 degrees of freedom.

Thus, it seems natural to identify this tensor with two spin-2
fields in the Fierz representation. In this way we obtain a
natural framework to connect spin-2 fields with the geometry. This
is our goal in the present paper\footnote{Note that we can choose
different forms of such restriction of torsion in order to limit
it to 20 independent components. In the next section we deal with
another choice.}.

Let us simplify here our analysis and consider the case in which
one of these tensors, say $B_{\mu\nu},$ vanishes. We postpone the
exam of the full case for the next section. We set
\begin{equation}
 2\, F_{\alpha\beta\mu} =  A_{\mu[\alpha,\beta]} +
F_{[\alpha}\eta_{\beta]\mu}, \label{0007bis}
\end{equation}

In order to analyse the curvature properties of this RCG let us remind that
the important quantity is not torsion itself but the so called
contortion, $K_{\alpha\beta\mu},$ defined by
\begin{equation}
\Gamma^{\epsilon}_{\mu\nu} = \{^{\epsilon}_{\mu\nu}\} +
K^{\epsilon}\mbox{}_{\mu\nu} \label{0008}
\end{equation}
In this section we restrict our analysis to the case in which the
riemannian geometry is flat. In other words, we can choose
conveniently the coordinate system in such a way that the
Christoffel symbol vanishes. This choice makes no restriction in
all calculations and all procedure throughout the whole paper is
completely covariant.

 Using equation (\ref{0002}) the contortion
is written as
\begin{equation}
K^{\epsilon}\mbox{}_{\mu\nu} = 2 F^{\epsilon}\mbox{}_{\nu\mu} +
F_{\nu}\, \delta ^{\epsilon}_{\mu} - F^{\epsilon} \, \eta_{\mu\nu}
\label{0009}
\end{equation}
or, in terms of the field $A_{\mu\nu}:$
\begin{equation}
K_{\epsilon\mu\nu} = A_{\epsilon\mu,\nu} - A_{\mu\nu,\epsilon},
\label{00010}
\end{equation}
from which follows the antisymmetry $K_{\epsilon\mu\nu} + K_{\nu\mu\epsilon} = 0.$

\subsection{Curvature tensor in RCG}

The curvature in an affine geometry is defined by
\begin{equation}
R^{\alpha}\mbox{}_{\sigma\beta\lambda} = \Gamma^{\alpha}_{\beta\sigma,\lambda}
- \Gamma^{\alpha}_{\lambda\sigma,\beta} +
\Gamma^{\alpha}_{\lambda\rho}\, \Gamma^{\rho}_{\beta\sigma} -
\Gamma^{\rho}_{\lambda\sigma} \, \Gamma^{\alpha}_{\beta\rho}.
\label{011}
\end{equation}
Since the metric tensor is Minkowskian this tensor contains only
the contributions that come from the torsion. The contracted
curvature tensor is then given by
 \begin{equation}
R_{\mu\nu} = K^{\alpha}\mbox{}_{\alpha\mu,\nu} - K^{\alpha}\mbox{}_{\nu\mu, \alpha}
+ K^{\alpha}\mbox{}_{\nu\rho}\, K^{\rho}\mbox{}_{\alpha\mu} -
K^{\alpha}\mbox{}_{\alpha\rho} \, K^{\rho}\mbox{}_{\nu\mu},
\label{012}
\end{equation}
Using equation (\ref{00010}) this curvature tensor can be
re-written as
\begin{equation}
R_{\mu\nu} = \Box A_{\mu\nu} - A^{\alpha}\mbox{}_{(\mu,
\nu)\alpha} + A_{,\mu\nu} + [KK]_{\mu\nu} \label{013}
\end{equation}
where
\begin{equation}
[KK]_{\mu\nu} \equiv K^{\alpha}\mbox{}_{\nu\rho}\,
K^{\rho}\mbox{}_{\alpha\mu} -  K^{\alpha}\mbox{}_{\alpha\rho} \,
K^{\rho}\mbox{}_{\nu\mu}. \label{013bis}
\end{equation}

Then using the decomposition in terms of the field $A_{\mu\nu}$ as
above this can be re-written as
\begin{eqnarray}
[KK]_{\mu\nu} &\equiv& \left(A_{\nu\alpha,\rho} -
A_{\nu\rho,\alpha}\right) \left(A^{\alpha\rho\mbox{}_,\mu} -
A_{\mu}\mbox{}^{\alpha,\rho}\right) \\ \nonumber &-&  \left(
A_{,\alpha} - A_{\alpha}\mbox{}^{\epsilon}
\mbox{}_{,\epsilon}\right) \eta^{\alpha\rho} \,
\left(A_{\nu\rho,\mu} - A_{\nu\mu, \rho}\right). \label{014}
\end{eqnarray}
From this expression it follows immediately that the trace
$[KK] \equiv [KK]_{\mu\nu} \, \eta^{\mu\nu},$ is
\begin{equation}
[KK] = -2\,U
\label{015}
\end{equation}
where $U$ is the invariant
\begin{equation}
U \equiv F_{\alpha\beta\mu}\, F^{\alpha\beta\mu} - F_{\alpha}\, F^{\alpha}.
\label{01511}
\end{equation}
Then, for the scalar of the curvature in RCG we obtain
\begin{equation}
R = 2\, \Box A - 2\, A^{\alpha\beta}\mbox{}_{,\alpha\beta} - 2\,U.
\label{016}
\end{equation}
or, using the equation (\ref{0007bis})
\begin{equation}
R = 2\,\left( F^{\alpha}\mbox{}_{,\alpha} - U \right)
\label{016bis}
\end{equation}

\subsection{Dynamics in RCG}
It seems natural to examine the dynamical torsion by choosing for
the Lagrangian function the scalar of curvature. From what we
have shown above this yields,
\begin{equation}
S = \int R \, d^{4}x =  -2 \, \int U \,d^{4}x
\label{017}
\end{equation}
up to a total divergence.

This dynamics is precisely the standard one proposed by Fierz
\cite{fierz} to describe a spin-2 field and corresponds to the
linear limit of Einstein General Relativity. Indeed, from the
above Lagrangian\footnote{See the appendix for further properties
of this description of a spin-2 field in terms of the three-index
tensor $F^{\alpha\beta\mu}$.} by varying $A_{\mu\nu}$ it follows:
\begin{equation}
\delta S = - \int 2 F^{\alpha(\mu\nu)}\mbox{}_{,\alpha}\,
 \delta A_{\mu\nu} \, d^{4}x
\label{018}
\end{equation}
Since we have the identity
\begin{equation}
F^{\alpha\mu\nu}\mbox{}_{,\alpha} = \frac{1}{2}
\, F^{\alpha(\mu\nu)}\mbox{}_{,\alpha} = - \,\frac{1}{2}\, G^{L}\mbox{}_{\mu\nu},
\label{01888}
\end{equation}
the equation of motion can be re-written as
\begin{equation}
\delta S = 2 \int G^{L}\mbox{}_{\mu\nu}\, \delta A^{\mu\nu} \,
d^{4}x, \label{018bis}
\end{equation}
where
\begin{equation}
G^{L}\mbox{}_{\mu\nu} \equiv \Box\,A_{\mu\nu} -
A^{\epsilon}{}_{(\mu,\nu)\, ,\epsilon} + A_{,\mu\nu} -
\eta_{\mu\nu}\, \left(\Box A - A^{\alpha\beta}{}_{, \alpha\beta}
\right), \label{019}
\end{equation}
where the label $L$ means the linear part of the Einstein equation
for GR\footnote{Note that there is a difference of factor 2
separating our definition of the operator $G^{L}\mbox{}_{\mu\nu}$
and the linear part of the Einstein operator $R_{\mu\nu} - 1/2 \,
R g_{\mu\nu}.$}. This accomplishes the proof of the following two
assertions:
\begin{itemize}
\item{Restricted torsion (10 degrees of freedom) describes a
symmetric second order tensor $A_{\mu\nu};$} \item{The dynamics
generated by the scalar of curvature (Hilbert-Einstein action)
yields the standard linear equation for the spin-2 field
$A_{\mu\nu}$.}
\end{itemize}

The next step is to go beyond this property to include other
degrees of freedom for torsion. We shall prove now that doubling
the number of variables from 10 to 20 causes the appearance of a
second spin-2 field.

\section{Cartan Geometry}

The generalization of the above formulation starts by the
modification of the representation of the torsion as in equation
(\ref{0002}) by the following one:
\begin{equation}
\tau^{\alpha}\mbox{}_{\mu\nu}  = F_{\mu\nu}\mbox{}^{\alpha} +
\frac{1}{2} g^{\alpha\epsilon}\mbox{}_{\mu\nu}\, A_{\epsilon} -
\eta_{\mu\nu}\mbox{}^{\alpha\epsilon} \, B_{\epsilon} \label{gc2}
\end{equation}
in which the trace and the pseudo-trace are given, respectively, by
\begin{equation}
\tau_{\alpha} = \frac{1}{2} \, F_{\alpha} = \frac{1}{2} \,
A_{\alpha}\label{gc3}
\end{equation}
\begin{equation}
\tau^{*}_{\alpha} = 4 B_{\alpha}. \label{gc4}
\end{equation}
The expression (\ref{gc2}) restricts the degrees of freedom of the
torsion only to 20 independent quantities. We limit all our
analysis to this case in order to deal only with two spin-2
fields. For the contortion, expression (\ref{0009}) becomes
\begin{equation}
K_{\epsilon\mu\nu} = 2 F_{\epsilon\nu\mu} +
g_{\epsilon\nu\mu\lambda} \, A^{\lambda} +
\eta_{\epsilon\nu\mu\lambda}\, B^{\lambda} \label{gc5}
\end{equation}
or
\begin{equation}
K_{\epsilon\mu\nu}=2A_{\epsilon\nu\mu}+
\eta_{\epsilon\nu}^{\rho\sigma} B_{\rho\sigma\mu}+
g_{\epsilon\nu\mu\lambda} \, A^{\lambda} +
\eta_{\epsilon\nu\mu\lambda}\, B^{\lambda} \label{gc5bis}
\end{equation}
 In this case we use the decomposition (\ref{gc6}) in terms of
two fields and $A_{\epsilon\nu\mu}$ and $B_{\epsilon\nu\mu}$ are
defined in (\ref{gc7}) and (\ref{gc8}).

The scalar of curvature is given by
\begin{equation}
R = K^{\alpha}\mbox{}_{\alpha\mu ,\nu} - K^{\alpha}\mbox{}_{\mu\nu ,\alpha}\, \eta^{\mu\nu}
+  [KK].
\label{gc10}
\end{equation}
Let us evaluate $[KK].$ From the above decomposition equation (\ref{gc5}) we find:
\begin{equation}
K^{\alpha}\mbox{}_{\alpha\mu} =  F_{\mu}
\label{gc11}
\end{equation}
\begin{equation}
K_{\mu\alpha\beta} \, \eta^{\alpha\beta} = - \,  F_{\mu}
\label{gc12}
\end{equation}

\begin{eqnarray}
K_{\alpha\mu\rho} \,K^{\rho\alpha\mu}  &=&  2 F_{\alpha\rho\mu}\,
\left(F^{\rho\mu\alpha}\, + F^{\mu\alpha\rho} \right)-
8 \, A^{\alpha} \, F_{\alpha} \nonumber \\
&+& 6 \,F_{\alpha}\, F^{\alpha} + 4 \, A_{\alpha}\, A^{\alpha} + 2
\,B_{\alpha} \, B^{\alpha} \label{gc13}
\end{eqnarray}

Collecting all this yields
\begin{eqnarray}
[KK] &=& - 2 B_{\alpha\beta\mu}\,B^{\alpha\beta\mu} +  2
B_{\alpha\beta\mu} \, B^{\alpha\beta\mu} \nonumber \\
&-&  2\, B_{\alpha}\, B^{\alpha} + 2 \,A_{\alpha} \, A^{\alpha}.
\label{gc15}
\end{eqnarray}
Then, for the scalar of curvature we find

\begin{equation}
R = 2 \,A^{\alpha}\mbox{}_{,\alpha} - 2 U[A_{\alpha\beta}] + 2
U[B_{\alpha\beta}]. \label{gc16}
\end{equation}
where the functional $U$ is given by equation (\ref{01511}), that
is
 $$U[A_{\alpha\beta}] \equiv A_{\alpha\beta\mu} \,
A^{\alpha\beta\mu} - A_{\alpha} A^{\alpha}, $$

and

$$ U[B_{\alpha\beta}] \equiv B_{\alpha\beta\mu}\,B^{\alpha\beta\mu} -
B_{\alpha} B^{\alpha}.$$

The important point to stress here concerns the relative signs of
$U[A_{\alpha\beta}]$ and $U[B_{\alpha\beta}],$ a property of
Cartan geometry in the two spin-2 representation.


Then the dynamics
\begin{equation}
S = \int R d^{4}x
\label{gc161}
\end{equation}
provides, for independent variation of the variables $A_{\mu\nu}$
and $B_{\mu\nu},$ equations of two independent non-interacting
spin-2 fields:
\begin{eqnarray}
G^{L}\mbox{}_{\mu\nu}(A) &\equiv& \Box\,A_{\mu\nu} -
A^{\epsilon}{}_{(\mu,\nu)\, ,\epsilon} + A_{,\mu\nu} \nonumber \\
&-& \eta_{\mu\nu}\, \left(\Box A - A^{\alpha\beta}{}_{,
\alpha\beta} \right) = 0. \label{gc162}
\end{eqnarray}
\begin{eqnarray}
G^{L}\mbox{}_{\mu\nu}(B) &\equiv& \Box\,B_{\mu\nu} -
B^{\epsilon}{}_{(\mu,\nu)\, ,\epsilon} + B_{,\mu\nu} \nonumber \\
&-& \eta_{\mu\nu}\, \left(\Box B - B^{\alpha\beta}{}_{,
\alpha\beta} \right) = 0. \label{gc1623}
\end{eqnarray}
It is worth to say that the curvature $R$ of the Cartan geometry,
within the framework we are developing here, yields linear
equations of motion for the associated spin-2 fields. This
property of linearity is typical for models in which  the torsion
is associated to fundamental fields. The association of spin-2 to
the metric, as in GR, leads inevitably to a nonlinear theory.

\section{Generalization to curved riemannian background}

All the above analysis was carried out in the case in which the
metric tensor has no Riemannian curvature. This means that the
metric tensor is identified with the Minkowski geometry. The
general case in which the Riemannian geometry is not flat is
straightforward and causes no difficulty. The attentive reader
could be concerned at this point with the known difficulty of
coupling spin-2 fields with an arbitrary Riemannian geometry. The
uses of the Fierz frame gives the answer to overcome such
difficulty as was shown recently \cite{novello-neves}. In this
case the total connection is given by
\begin{equation}
\Gamma^{\alpha}_{\mu\nu} = \{{}^{\alpha}_{\mu\nu}\} +
K^{\alpha}_{\mu\nu} \protect\label{r1}
\end{equation}
The Christoffel symbol  $\{{}^{\alpha}_{\mu\nu}\}$  is constructed
in terms of the metric tensor $g_{\mu\nu}$ and its derivatives;
$K^{\alpha}_{\mu\nu}$   is the contortion defined previously. The
curvature tensor contains an additional term:
\begin{equation}
R_{\mu\nu} = \hat{R}_{\mu\nu}  + K^{\alpha}\mbox{}_{\alpha\mu;\nu}
- K^{\alpha}\mbox{}_{\nu\mu ; \alpha} +
K^{\alpha}\mbox{}_{\nu\rho}\, K^{\rho}\mbox{}_{\alpha\mu} -
K^{\alpha}\mbox{}_{\alpha\rho} \, K^{\rho}\mbox{}_{\nu\mu},
\label{r2}
\end{equation}
where $\hat{R}_{\mu\nu}$  is the contracted curvature tensor of
the riemannian sector of the curvature. The covariant derivative
$;$ must be taken in this sector. The complete dynamics in such
general case is provided by the scalar $R$:
\begin{equation}
R = \hat{R} + 2 \,A^{\alpha}\mbox{}_{;\alpha} - 2
U[A_{\alpha\beta}] + 2 U[B_{\alpha\beta}]. \label{r3}
\end{equation}
We postpone the complete analysis of  this general case for
another paper.

\section{The interaction with matter}

The fundamental constituents of matter are associated to
elementary particles that have non integer spins like quarks and
leptons. This property will lead us to concentrate here in the
analysis of the coupling of spin-2 fields $A_{\mu\nu}$ and
$B_{\mu\nu}$ to spinors $\Psi(x).$

In order to understand the coupling of the matter field $\Psi$
with the spin-2 displayed in torsion let us review briefly the
similar situation of coupling  such a field $\Psi$ with gravity.
In other words let us analyse the coupling of spin-2 to matter in
the Einstein-frame and in the Fierz-frame.

\subsection{The gravitational interaction of the fermionic field}

 In the standard minimal coupling framework the
interaction of matter to the spin-2 fields, within the
geometrization scheme that we presented in the previous chapters,
consists simply in the operation of analysis of Dirac equation of
motion for a spinor embedded in the modified geometry: a
Riemannian one in case of gravity (the Einstein frame) and a
Cartan geometry in case of the extra two spin-2 fields (the Fierz
frame). This is made by the introduction of a covariant derivative
of the spinor. In this section we deal only with the Riemann
sector. We set

\begin{equation}
\nabla_{\mu} \, \Psi \equiv \partial_{\mu} \Psi - \Gamma_{\mu}
\Psi \label{iwm1}
\end{equation}
where the Fock-Ivanenko spinor connection is given in the standard
way:
\begin{equation}
\Gamma_{\mu} = -\, \frac{1}{4} \, \left( e^{\lambda}_{b}
\partial_{\mu}e^{a}_{\lambda} -  \{{}^{a}_{b\,\mu}\} \, \right)
\sigma^{b}{}_{a}. \label{iwm2}
\end{equation}
and
\begin{equation}
\sigma^{b}_{a} \equiv \frac{1}{2} \, ( \gamma^{b} \gamma_{a} -
\gamma_{a} \gamma^{b} )
 \label{iwm2bis}
\end{equation}

In this expression  $e^{\lambda}_{b}$ is a set of orthogonal
tetrads and
\begin{equation}
\{{}^{a}_{b \,\mu}\}  \equiv \{{}^{\alpha}_{\beta\,\mu}\} \,
 e_{\alpha}^{a} \, e^{\beta}_{b},
\label{iwm3}
\end{equation}

The uses of the minimal coupling principle yields for the action
of the spinor the form

\begin{eqnarray}
S &=& \frac{i}{2} \,\int \sqrt{-g} \, ( \bar{\Psi}
\gamma^{\alpha}\, \nabla_{\alpha}\, \Psi - \bar{\nabla}_{\alpha}
\bar{\Psi} \, \gamma^{\alpha}\,\Psi )d^{4}x \nonumber \\
&-& \,\int \sqrt{-g}\, \mu^{2} \,\bar{\Psi}\Psi d^{4}x
\label{iwm4}
\end{eqnarray}

In other words the minimal coupling principle yields for the
gravitational interacting action:
\begin{equation}
S_{int} =  \int \sqrt{-g} \, T_{\mu\nu} g^{\mu\nu}
d^{4}x\label{g1}
\end{equation}
where the energy-momentum tensor is given by
\begin{equation}
T_{\mu\nu} = \frac{i}{4} \, \{ \bar{\Psi} \, \gamma_{(\mu}
 \nabla_{\nu)} \Psi - \nabla_{(\mu} \bar{\Psi} \,
\gamma_{\nu)} \,\Psi \}  - \frac{\mu^{2}}{4} \, \bar{\Psi} \Psi
\,g_{\mu\nu} \label{g2}
\end{equation}

\subsection{The interaction of the fields $ A_{\mu\nu}$ and
$B_{\mu\nu}$ with the fermions}

We will follow a similar procedure in order to describe the
coupling of the fermions with the extra two spin-2 fields
$A_{\mu\nu}$ and $B_{\mu\nu}$ represented in the Fierz frame. This
means that we will make the analysis of the Dirac equation of
motion for a spinor embedded in the Cartan geometry.

The covariant derivative is written as above
\begin{equation}
\nabla_{\mu} \, \Psi \equiv \partial_{\mu} \Psi - \Gamma_{\mu}
\Psi \label{iwm1bis}
\end{equation}
where now the Fock-Ivanenko spinor connection is given in terms of
the affine conection $\Gamma^{\alpha}_{\beta\, \mu}$ given by
(\ref{0008}) and (\ref{gc5}). The action takes the same form as in
equation (\ref{iwm4}).

Noting that
\begin{equation}
\gamma^{5} \equiv - \frac{1}{4!}\eta^{\mu\nu\alpha\lambda}\,
\gamma_{\mu} \gamma_{\nu} \gamma_{\alpha} \gamma_{\lambda},
\label{iwm4bis}
\end{equation}
and using the above connection we can re-write the action under
the form

\begin{eqnarray}
S &=& i \,\int \sqrt{-g} ( \bar{\Psi} \gamma^{\alpha}\,
\nabla_{\alpha}\, \Psi
 + \frac{1}{4} \, K_{\mu\nu\alpha} \, \eta^{\mu\nu\alpha\lambda}
 \, \bar{\Psi} \, \gamma_{\alpha} \gamma^{5} \,\Psi ) d^{4}x \nonumber \\
 &-&  \,\int \sqrt{-g}\mu^{2} \,\bar{\Psi}\Psi d^{4}x
\label{iwm5}
\end{eqnarray}

From the expression (\ref{gc5}) of the contortion we have

\begin{equation}
K_{\mu\nu\alpha} \, \eta^{\mu\nu\alpha\lambda} = \left( 2
F_{\mu\alpha\nu} + g_{\mu\alpha\nu\lambda} B^{\lambda} -
\eta^{\mu\nu\alpha\lambda} B^{\lambda} \right)
\eta^{\mu\nu\alpha\lambda} \label{iwm6}
\end{equation}
which reduces to

\begin{equation}
K_{\mu\nu\alpha} \eta^{\mu\nu\alpha\lambda} = 2 B^{\lambda}
\label{iwm7}
\end{equation}

Thus the action assumes the  form

\begin{eqnarray}
S &=& \int \sqrt{-g}( i \bar{\Psi} \gamma^{\alpha}\,
\nabla_{\alpha}\, \Psi -  \mu^{2} \,\bar{\Psi}\Psi d^{4}x
\nonumber \\
&+& \frac{1}{2} \, \int \sqrt{-g}   \, \bar{\Psi}
  \, \gamma_{\lambda} \gamma^{5} \,\Psi \, B^{\lambda} \,d^{4}x
\label{iwm8}
\end{eqnarray}

which shows that matter minimally coupled to Cartan background
geometry interacts only with the $B_{\mu\nu}$ field. Let us
analyse a little more carefully such interaction in order to
understand the differences of the matter coupling to the spin-2
field which defines the metric tensor in the case of General
Relativity.

\subsection{The interaction of spin-2 fields with the fermionic
matter}

The minimal coupling principle of the matter field $\Psi$ with
torsion yields the interacting term  (see equation (\ref{iwm8}))
\begin{equation}
S_{int} = \frac{i}{2}\,\int \sqrt{-g}\, B^{\lambda} \, \bar{\Psi}
\, \gamma_{\lambda} \gamma^{5} \,\Psi d^{4}x \label{iwm9}
\end{equation}
in which we used the decomposition of torsion  (\ref{gc2}). The
quantity  $B^{\lambda}$ is given in terms of the field
$B_{\mu\nu}$ as

\begin{equation}
B_{\lambda} =   \nabla_{\mu} \, (B^{\alpha\beta} \, \,
g_{\alpha\beta} \delta^{\mu}_{\lambda} - B_{\lambda}^{\mu})
\label{iwm11}
\end{equation}
The interacting term can be re-written, up to a total divergence,
as

\begin{equation}
S_{int} =  \int \sqrt{-g} \,  \{B^{ \mu\nu} - B \, g^{\mu\nu} \}
\, J_{\mu\nu} d^{4}x
  \label{iwm12}
\end{equation}
where
\begin{equation}
J_{\mu\nu} =  \nabla_{\mu} (\bar{\Psi} \, \gamma_{\nu} \gamma^{5}
\,\Psi )
\label{iwm13}
\end{equation}

The crucial distinction from this coupling (\ref{iwm12}) to the
previous one -(\ref{g1})concerns the behavior of the current
$J_{\mu\nu}$ and $T_{\mu\nu}$ under the operation of hermiticity.
The energy-momentum tensor is written as the product of the
imaginary quantity $i$ with the difference of two terms containing
the derivative operator. Instead of the imaginary quantity, the
presence of the hermitian operator $\gamma^{5}$ in the current
$J_{\mu\nu}$ is precisely the condition that allows such quantity
to be written as a total derivative. This is the crucial point
that makes the distinction between these two forms of coupling. In
the case of gravity the corresponding term
\begin{equation}
S_{int} =  \int \sqrt{-g} \, J_{\mu\nu} g^{\mu\nu} d^{4}x
\label{g2bis}
\end{equation}
is a total divergence and thus does not contribute to the
dynamics. This is in agreement with the result that gravity is a
parity conserving force.

\section{Conclusion and some comments}
The purpose of the present paper was to show the existence of two
non-equivalent ways to incorporate spin-2 fields into the
geometric structure of 4-dimensional spacetime. These modes are
related to two distinct representations of spin-2 fields. One of
them, the Einstein-frame, uses a symmetric second order tensor
$\varphi_{\mu\nu}$ to represent the field; the other, the
Fierz-frame, uses a third rank anti-symetric tensor
$F_{\alpha\beta\mu}.$ In the first case, the natural way to
geometrize the field consists in the association of
$\varphi_{\mu\nu}$ to the metric tensor, as we showed in equation
(\ref{001}); in the second case the natural way is the association
of the Fierz tensor to the torsion as we did in equation
(\ref{gc2}). The most important difference between these two
representations appears in their formal implementation. In the
case of the Einstein-frame one deals with a non-linear structure.
This is evident from the expression (\ref{002}) of the inverse
metric. In the Fierz representation instead all the process of
geometrization may be linearly implemented. This property led to
the choice of the Einstein-frame for the geometrization of the
spin-2 field associated to the gravity. All remaining spin-2
fields may be described in the Fierz representation. Besides, in
the Fierz frame the traditional ambiguity that was present in
coupling higher spin field to gravity is overcomed
\cite{novello/neves}. Let us point out the property which follows
from using the associated Hilbert-Einstein action based on the
scalar (\ref{r3}). Although all this structure has a geometrical
meaning, if one interprets the extra two spin-2 fields in the
realm of Einstein GR as matter fields, then a curious property
appears. The equation of motion for each of the extra spin-2
fields is linear and the energy contribution has opposite signs.
This allow to recognize the possibility of the existence of a
situation in which the energies of both fields cancel out
generating a sort of {\it torsion vacuum state.}

Finally, a remarkable property of such a description appears when
we take into account its corresponding interaction with matter.
Only one of the fields $B_{\mu\nu}$ has a direct, minimal,
coupling with the fermionic field. Besides, such interaction
violates parity. This suggests a possible existence of an
interaction of this spin-2 field with the matter that violates
parity. This is under investigation.

\section{Ackowledgement}
I would like to thank the participants of the {\it Pequeno
Semin\'ario} of CBPF and specially Dr. J. M. Salim for comments and discussions.
\end{document}